\documentclass[12pt]{iopart} 

\usepackage[]{graphicx}  
\newcommand{\sinc}{{\rm sinc}}
\begin{document}

\title[Two-colour lasers with optical injection and feedback]{Wavelength switching dynamics of two-colour semiconductor lasers with optical injection and feedback}

\author{S. Osborne$^1$, P. Heinricht$^{1,2}$, N. Brandonisio$^{1,2}$, A. Amann$^{1,3}$, S. O'Brien$^1$}
\address{$^1$ Tyndall National Institute, University College Cork, Lee Maltings, Cork, Ireland}
\address{$^2$ Department of Physics, University College Cork, Ireland}
\address{$^3$ School of Mathematical Sciences, University College Cork, Ireland}

\ead{simon.osborne@tyndall.ie}

\begin{abstract}
The wavelength switching dynamics of two-colour semiconductor lasers with optical injection and feedback are presented. These devices incorporate slotted regions etched into the laser ridge waveguide for tailoring the output spectrum. Experimental measurements are presented demonstrating that optical injection in one or both modes of these devices can induce wavelength bistability. Measured switching dynamics with modulated optical injection are shown to be in excellent agreement with numerical simulations based on a simple rate equation model. We also demonstrate experimentally that time-delayed optical feedback can induce wavelength bistability for short external cavity lengths. Numerical simulations indicate that this two-colour optical feedback system can provide fast optical memory functionality based on injected optical pulses without the need for an external holding beam.
\end{abstract}

\submitto{\SST}
\maketitle 

\section{Introduction}

Semiconductor Fabry-Perot (FP) lasers have the simplest cavity geometry as the resonator can be formed between two mirrors defined by the cleaved facets. However, despite being very convenient to manufacture, simple FP devices are unsuitable for many applications because of their characteristic multimode lasing spectrum. There are alternative cavity geometries that can result in single mode emission such as the distributed feedback laser, where a Bragg grating adjacent to the active region provides feedback. These devices are commonly deployed in optical communication systems but they have the disadvantage that typically two or more epitaxial growth steps are required to fabricate the device \cite{coldren}.

An alternative technique is to incorporate a number of scattering centres into a FP laser that can modify the lasing spectrum. This technique may require no additional processing where scattering centres in the form of slots are introduced at the same time as the ridge waveguide itself is formed \cite{corbett_1995}. Although the position of the slots along the device can be found empirically or with use of a generic algorithm, we have shown that the device geometry can be determined using an inverse problem solution based on a perturbative calculation of the threshold gain of the longitudinal modes \cite{obrien_2005}. As the device geometry can now be obtained from simple analytic expressions, this technique can be extended to design devices with a predetermined number and spacing of lasing modes. The wavelength switching dynamics of two-colour devices designed in this way are the focus of this paper.

Semiconductor lasers with a fixed and predetermined number and spacing of primary modes are of interest for a number of applications. For example, two-colour devices that oscillate simultaneously on both modes are useful for terahertz generation by photomixing \cite{tani_2005}. Another interesting prospect is fast switching between discrete wavelength channels to enable optical signal routing \cite{white_2002, brandonisio_2011}. Switching between wavelength channels can also be the basis for all-optical signal processing based on optical memory and logic functions \cite{papadimitriou_2003, dorren_2003}.

The structure of this paper is as follows. In section \ref{sect:twocolour}, we describe our design approach and present measured characteristics of a free running two-colour device. In section \ref{sect:memories}, we first show that both single and dual injection can induce wavelength bistability and we demonstrate an all-optical memory function in each case. A simple rate equation model for the two-colour laser is introduced which provides excellent agreement with experimental results. Finally, we discuss our results obtained with time delayed optical feedback and show that wavelength bistability can be induced in the absence of an external holding beam. Numerical simulations based on an extended rate equation model are again in excellent agreement with experiment and suggest that injected optical pulses can enable switching at GHz rates between single-mode states of the self-sustained feedback system.

\section{Device design and free running lasing characteristics}\label{sect:twocolour}

The discussed two-colour lasers are edge-emitting devices with slotted features etched into the laser ridge waveguide. By treating these slots as index step features along the cavity, a transmission matrix calculation can directly relate the threshold gain modulation in wave-number space to the refractive index profile along the device. With a refractive index step, $\Delta n$, and number of features, $N$, a first order scattering approximation describes the coupling between the external mirrors and the features, but neglects the coupling between the features themselves. A set of self-consistent equations for lasing mode thresholds and frequencies in the perturbed cavity can then be derived \cite{osborne_2007}. 

We select a given longitudinal mode, $m_{0}$, as our origin in wavenumber space and we assume quarter wavelength features with respect to this mode. Considerable simplification results if we assume that half-wave and quarter-wave sub-cavities are formed to the left and right of each feature. In this case the index profile forms a sampled Bragg grating and the threshold modulation is an even function. We find that the threshold gain, $G_{th}$, of longitudinal mode $m$ is given by
\begin{eqnarray}
G_{th}\left(m\right)\! &\! = &\! G_{th}^{FP}\!+\!\frac{\Delta n}{nL_c\sqrt{}r_1r_2}\cos\left(m_0\pi\right)\cos\left(\Delta m\pi\right) 
\nonumber \\
 & &\times \sum_{j=1}^{N}A\left(\epsilon_j\right)\sin\left(2\pi\epsilon_j m_0\right)\cos\left(2\pi\epsilon_j\Delta n\right)
\end{eqnarray}
where
\begin{equation}
A\left(\epsilon_j\right)\!=
\!\vert r_1 \vert\exp\left(\epsilon_jL_cG_{th}^{FP}\right)\!-\!\vert r_2 \vert\exp\left(-\epsilon_jL_cG_{th}^{FP}\right).
\end{equation}
In the above, $G_{th}^{FP}$ is the threshold gain of the unperturbed FP cavity of length $L_c$ and $\epsilon_j$ is the fractional position of the centre of each feature measured from the centre of the cavity. The refractive index is denoted by $n$, and the reflectivities of the mirrors are $r_1$ and $r_2$. The function $A\left(\epsilon_j\right)$ is present because the change in threshold gain is proportional to the difference in round-trip gain to the left and right of each feature. The threshold gain is expressed with respect to the longitudinal mode index $m=m_0+\Delta m$, thereby enabling us to use Fourier analysis to build up a particular threshold gain in wave-number space. 

\begin{figure}
  \begin{center}
   \includegraphics[width=\columnwidth]{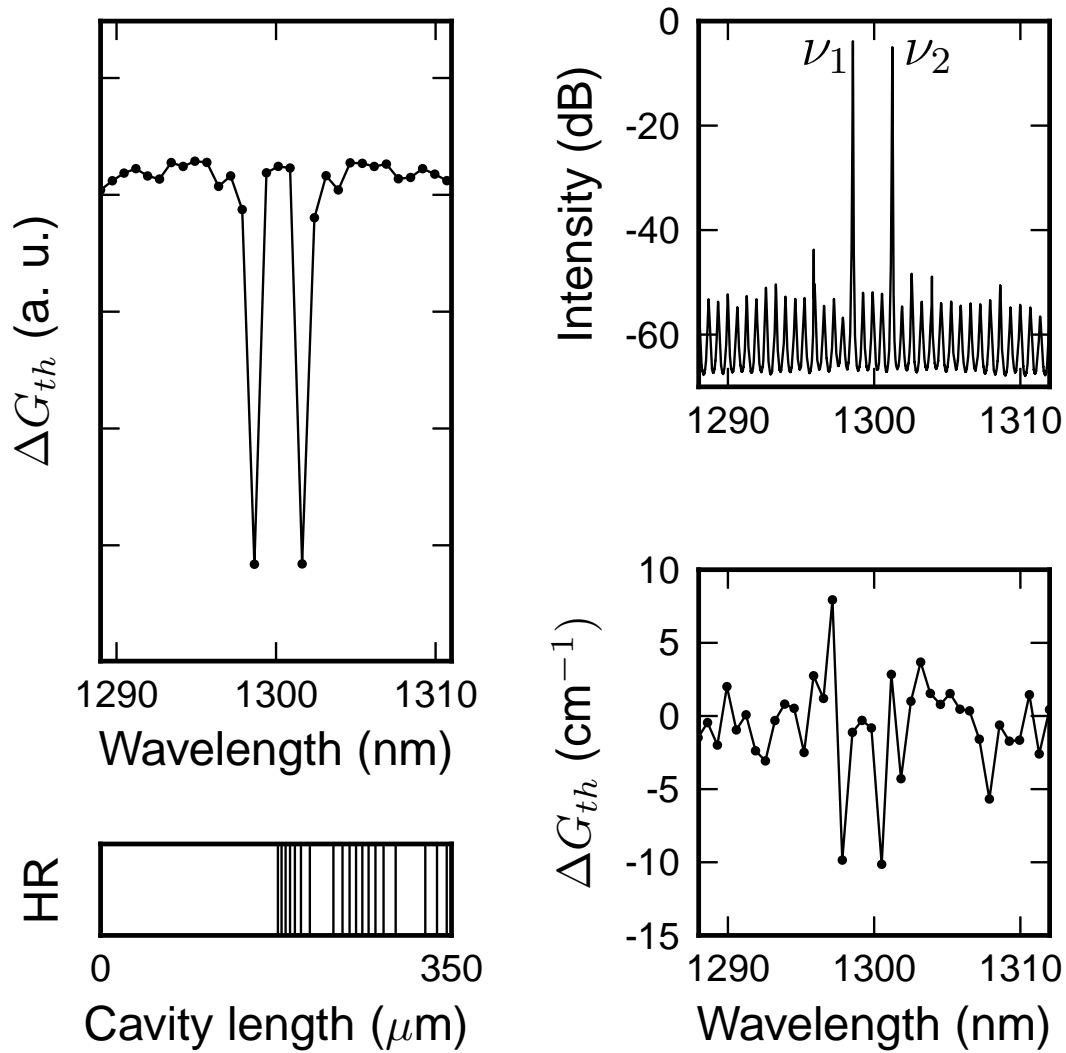}
   \caption{\label{fig:geometry} The left panel shows the ideal threshold gain spectrum for the two colour laser. The bottom left panel shows the position of the slotted regions and the left mirror has a high reflective [HR] coating. The upper right panel shows the optical spectrum of the two-colour laser at 43 mA. In the lower right panel, the measured change in threshold gain shows the selection of the two primary modes.}
  \end{center}
\end{figure}

For the two colour laser, a pair of longitudinal modes should experience a reduced threshold gain while leaving the others unperturbed. In this case the ideal threshold gain modulation comprises two sinc functions with spacing $a$ centred at mode $m_0$ 
\begin{equation}
G_{th}-G_{th}^{FP}\sim\sinc\left(\Delta m +a/2\right)+\sinc\left(\Delta m -a/2\right).
\end{equation}
This function has a Fourier transform $\Pi\left(\epsilon\right)\cos\left(\pi a\epsilon\right)$, where $\Pi\left(\epsilon\right)$ is the unit rectangle centred at $\epsilon=0$. The distribution of features is found by discretely sampling the product of this Fourier transform with the function $\lbrack A\left(\epsilon_j\right)\rbrack^{-1}$. As this feature density changes sign along the device length, $\pi/2$ phase shifts are introduced into the index pattern at these positions. Figure \ref{fig:geometry} shows a calculation of the modal threshold gain for a device of length 350 $\mu\mbox{m}$ with emission centered near 1300 nm. The primary mode spacing is four FP modes, which corresponds to 480 GHz. The device geometry is shown in the lower left panel of figure \ref{fig:geometry} where a total of 19 slots were introduced. 

In this paper we present results obtained with two different two-colour FP devices. The first is based on the design shown in figure \ref{fig:geometry}. The second is a 545 $\mu m$ long device designed to emit around 1550 nm. This device has a primary mode spacing of six FP modes which also corresponds to a spacing of 480 GHz. Both devices are fabricated using InP-based epitaxial structures and were characterized in a temperature controlled environment.

The optical spectrum of the 1300 nm device is shown in figure \ref{fig:geometry} at a device current of 43 mA. At this current value the two primary modes oscillate simultaneously with approximately equal intensities. The mode selectivity due to the introduction of the slots can be quantified using the Hakki-Paoli technique where the resultant net modal gain includes contributions from the modal gain of the active region and the change due to the additional features. The change in the threshold gain shown in the bottom right panel of figure \ref{fig:geometry} is found by subtracting the material modal gain from the measured modal gain and shows a $\sim$10 $\mbox{cm}^{-1}$ reduction in gain of the the two primary modes compared with the background longitudinal modes. 

Simultaneous lasing of the primary modes of the device shown indicates that these modes are weakly coupled. For devices with frequency spacings less than approximately 300 GHz, strong coupling sets in and simultaneous lasing is not possible. Instead, mode hopping between co-existing single mode states is observed over a small hysteresis region \cite{obrien_2010}. This wavelength bistability occurs in quantum well lasers when the gain cross-saturation exceeds the self-saturation, indicating that the frequency spacing is now less than the homogeneous linewidth of the gain material. Although noise driven switching is undesirable for many applications of two-colour devices with spacings of order 100 GHz, our recent work has shown that the introduction of a second pair of modes can circumvent this limitation by enabling mode-locking of the device \cite{obrien_2010}.

In the next section we introduce our model of the multimode semiconductor laser, which inlcudes a single averaged carrier density variable and a general cross and self saturation of the gain. In this way we explicitly neglect the coupling effect due to the carrier density grating that forms in the presence of two lasing modes. This dynamic grating leads to an asymmetric interaction with enhanced gain at the long wavelength mode, and has been shown to play a role in the free running switching dynamics of Fabry-Perot lasers \cite{yacomotti_2004}. However, we neglect this term on account of the very large separation of the primary modes of the device (480 GHz), which suggests that the dynamic grating is weakly developed.

\section{Wavelength switching dynamics with optical injection and feedback}
\label{sect:memories}

All-optical signal processing has been the subject of extensive research in the last decade and a fundamental function required for information processing is memory operation. Several different solutions have been proposed in recent years for the implementation of all-optical memories and many approaches to this problem rely on the co-existence of lasing modes of a given system with injected optical pulses for switching between these states. There are a number of potential optical memory systems, those based on counter-propagating modes in micro-ring devices\cite{Liu_2010}, coupled ring lasers\cite{hill_2004} and those using the two polarization modes in vertical surface emitting devices \cite{mori_2006}, which can all have switching times less than 100 ps.  

Although switching performance in conventional edge-emitting diode lasers can be constrained by speed limitations and power consumption, these devices represent testbeds for the study of the underlying dynamics based on generic systems of equations. Dynamical behaviors found in conventional diode laser systems may be scalable to integrated devices based on photonic crystals or microcavities. Recent examples include wavelength bistability induced by gain cross saturation in a micro-cavity laser \cite{zhukovsky_2009}, and injection locking bistability in a photonic crystal laser that supports two lasing modes \cite{chen_2011} where the latter has switching times of 60 ps. Systems such as these promise to overcome limitations of size, speed and power consumption associated with conventional devices.

\subsection{Single optical injection and rate equation model}

\begin{figure}
 \begin{center}
  \includegraphics[width=\columnwidth]{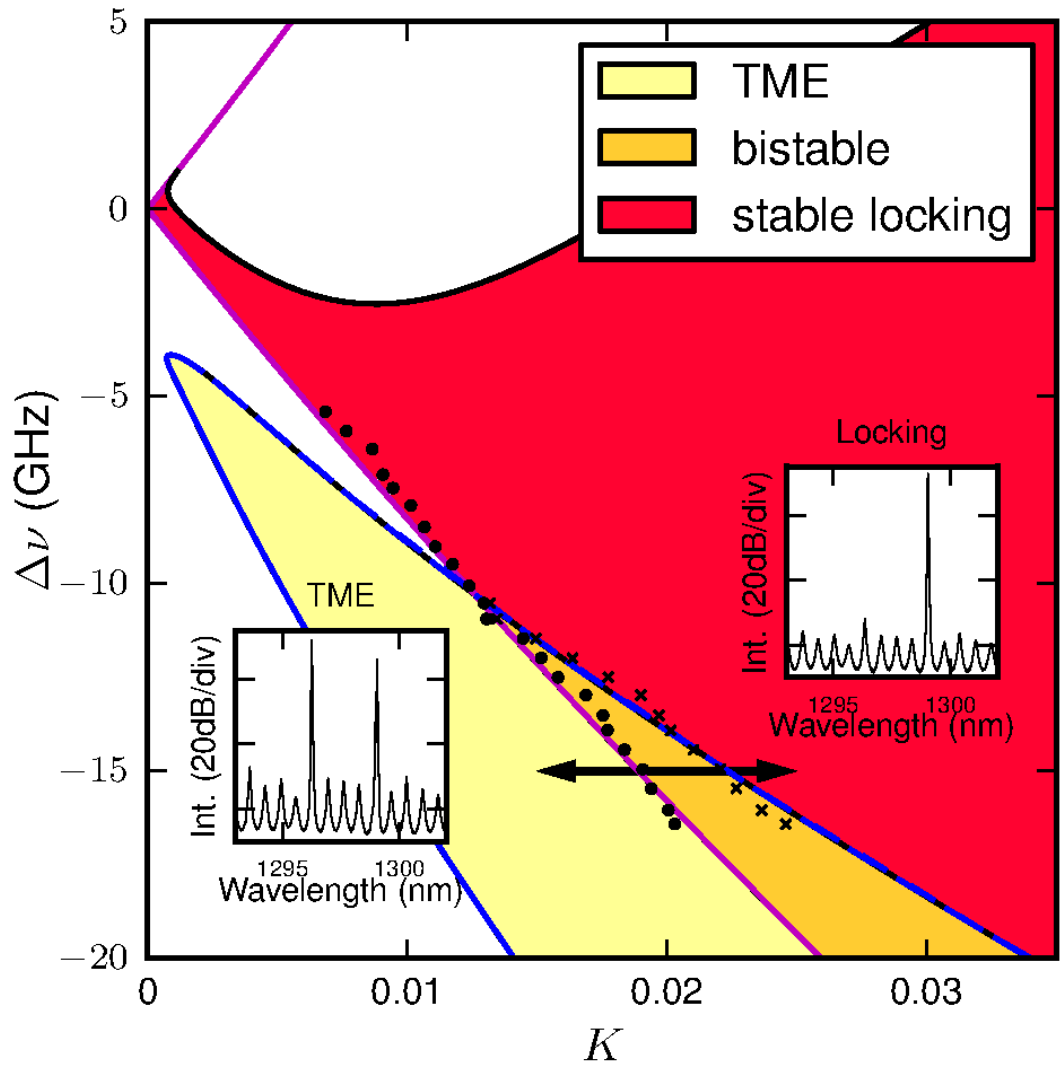}
   \caption{\label{fig:single_bif} The bifurcation diagram in the $\nu-K$ plane. The light to dark (yellow to red in color) shaded areas show the regions relevant to the all-optical memory: two-mode equilibrium (TME), bistability and the stable locking, respectively. The points show the experimentally measured locking ($\times$) and unlocking ($\bullet$) boundaries. The two insets show the optical spectra of the two bistable states TME (left) and stable locking (right).  }
  \end{center}
\end{figure}

As our first example we consider a two-colour laser subject to optical injection in one of the primary modes of the device. In the insets of figure \ref{fig:single_bif} we show the co-existing states associated with the injection locking bistability of the laser. One of these states is the conventional single mode locked state of the system, while the other is a two-mode steady state. One can see that the device may operate as a memory element where the intensity of the uninjected mode can be switched between an ``off'' state and an ``on'' lasing state with a contrast ratio greater than 35 dB \cite{osborne_2009}. 
 
To understand the structure of the bifurcations that lead to this bistability we have adapted the well known model of a single mode laser with optical injection to account for the presence of a second lasing mode \cite{varangis_1997}. If we allow for the general case of optical injection and feedback in each mode of the laser, the system of equations in normalized units may be written as follows \cite{osborne_2009_PRA,brandonisio_2012}:

\setlength{\arraycolsep}{0.0em}
\begin{eqnarray}
\label{eq:sys}
&& \dot{e}_1=(1/2)(1+i\alpha)(g_1 (2n+1) -1) e_1 
+ e_1^{\mbox{\tiny D}} + e_1^{\mbox{\tiny inj}} \nonumber\\
&& \dot{e}_2=(1/2)(1+i\alpha)(g_2 (2n+1) -1)e_2
+ e_2^{\mbox{\tiny D}} + e_2^{\mbox{\tiny inj}} \\
&& T\dot{n}= p-n-(2n+1)(g_1 |e_1|^2 + g_2 |e_2|^2). \nonumber
\end{eqnarray}
\setlength{\arraycolsep}{5pt}

In the above equations, $e_m$ are the normalized complex electric fields associated with the two modes of the laser and $n$ is the normalized excess carrier density in the cavity. Optical injection is included in the normalized complex fields $e_m^{\mbox{\tiny inj}} = K_m\exp(i\Delta\omega_m^{\mbox{\tiny inj}}t)$ where $K_m$ is the injected field strength. Here the normalized detunings $\Delta\omega_m^{\mbox{\tiny inj}}$ are measured from the free running lasing mode frequencies and for clarity we refer to the physical detunings $\Delta\nu_m$. $e_m^{\mbox{\tiny D}} = \eta e_m(t-\tau)\exp(-i\varphi_m)$ are the normalized complex fields delayed by the external cavity round trip time $\tau$. The feedback strength is $\eta$ and $\varphi_m$ is defined such that $\varphi_1 = \omega_1 \tau = \varphi$ and $\varphi_2 = \omega_2 \tau = \omega_1 \tau + \Delta\omega \tau = \varphi + \Delta\varphi$, where $\omega_1$ and $\omega_2$ are the free running lasing mode frequencies and  $\Delta\omega = \omega_2 - \omega_1$. $\varphi$ and $\Delta\varphi$ are the reference feedback phase and the feedback phase difference of the modes.  

The quantity $g_m = g_m^0(1+\epsilon \sum_j \beta_{mj}|e_j|^2)^{-1}$ is the nonlinear gain function, where $g_m^0$ is the linear gain, $\epsilon \beta_{mm}$ is the gain self saturation and $\epsilon \beta_{mj}$ is the gain cross saturation. The phase-amplitude coupling is given by $\alpha$, $p$ is the normalized pump current, while $T$ is the product of the cavity decay rate $\gamma$ with the carrier lifetime $\tau_s$. In our simulations we choose the following parameters $\alpha=2.6$, $T=800$, $\gamma=9.8\times 10^{11} \mbox{s}^{-1}$, $g_m^0=1$, and $\epsilon=0.01$. In order to be consistent with the existence of two-mode solutions in the free running laser, we have taken $\beta_{12}=\beta_{21}=2/3$ and $\beta_{11}=\beta_{22}=1$. The parameter values chosen have provided excellent agreement with the results of optical injection experiments with two-colour devices \cite{osborne_2009, osborne_2009_PRA}.

\begin{figure}
 \begin{center}
  \includegraphics[width=\columnwidth]{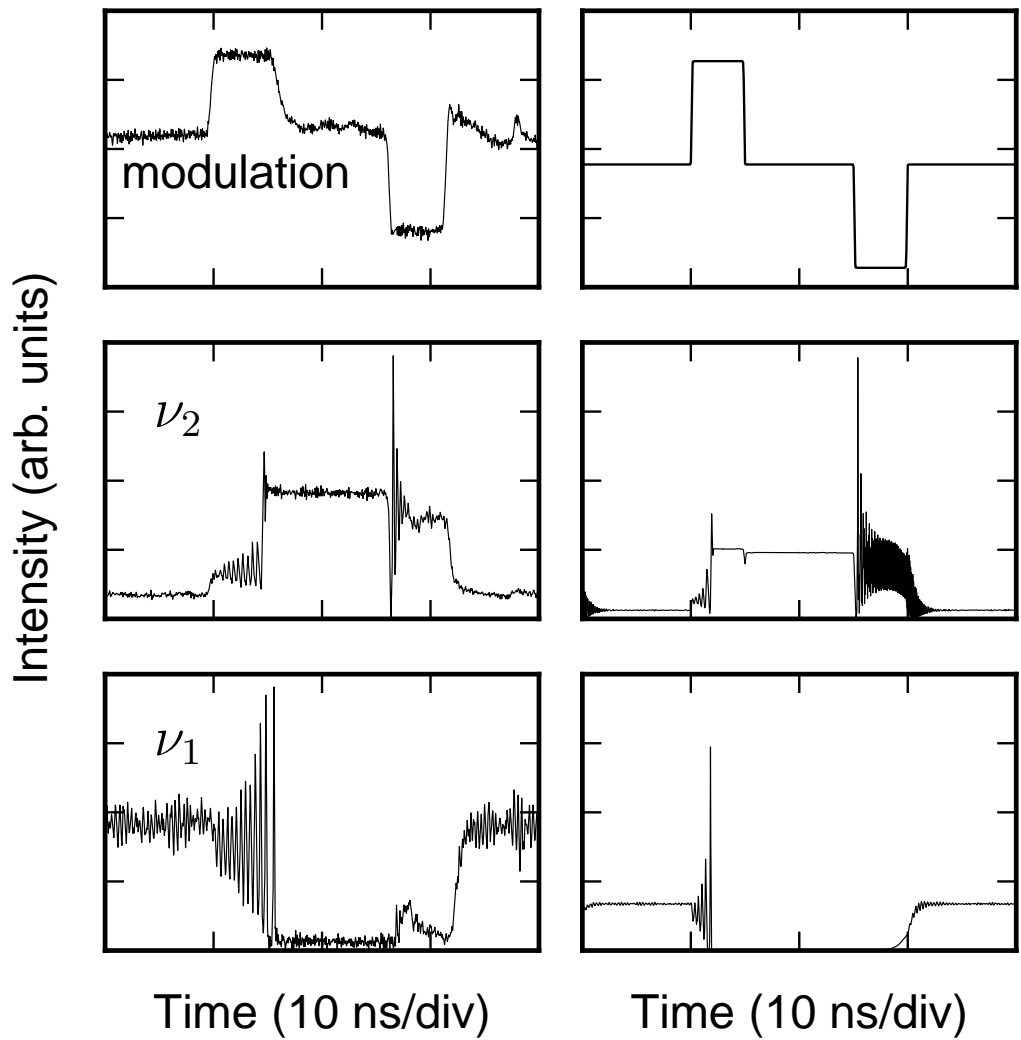}
   \caption{\label{fig:single_switch} The experimental (left) and calculated (right) time series of the injected mode ($\nu_2$) and uninjected mode ($\nu_1$) for the modulation shown and $\nu=-14 \mbox{GHz}$. }
  \end{center}
\end{figure}

For the case of single injection in the long wavelength mode of the laser we set $e_1^{\mbox{\tiny D}} = e_1^{\mbox{\tiny inj}} = e_2^{\mbox{\tiny D}} = 0$. Note that in this case the phase of $E_{1}$ is decoupled in the above system of equations, reflecting the fact that the primary mode spacing is in the highly non-degenerate regime. The bifurcation structure that governs the injection locking bistability in the two-colour device can now be understood from the global bifurcation diagram in the $\Delta \nu$ vs. $K$ plane that is shown in figure \ref{fig:single_switch}. These curves were calculated using the numerical continuation tool AUTO \cite{auto}. The light to dark (yellow to red in colour) shaded regions of figure \ref{fig:single_bif} show the respective regions of two-mode steady or equilibrium states [TME], bistability and single-mode locking. Experimentally, we identify the transitions between locking and unlocking shown as the points in figure \ref{fig:single_bif}. The experimental data shows very good agreement with the model where bistability appears at $\Delta\nu<-10$ GHz. Similar bifurcation structures are also seen in other two-mode lasers, vertical cavity surface emitting lasers \cite{gatare_2009} and ring lasers \cite{coomans_2010}.

An all-optical memory function based on the injection locking bistability can be achieved by appropriately modulating the injected field, as indicated by the arrow in figure \ref{fig:single_bif} for a frequency detuning of $\Delta\nu=-14\mbox{ GHz}$. The measured time series of the modulated injected signal and individual modes are shown in the left of figure \ref{fig:single_switch}. The modulated optical injection (top) shows set and reset inputs of 5 ns duration with rise times less than 100 ps. The right of figure \ref{fig:single_switch} shows the calculated time series of the all-optical memory and we see very good agreement with the experimental results. 

\subsection{Dual Optical Injection}

In the case of single optical injection, we can regard the injected mode of the laser as the writing channel with readout of the memory state from the uninjected mode. While this configuration may be appropriate for certain applications, one consequence is the slow switching speed of the system when the single-mode state unlocks. The reason for this limitation is the fact that in order to switch from the injection locked state to the two-colour state, the injected signal is modulated close to zero and thus the device switches as an essentially free running laser.  

In order to obtain faster switching, one approach is to employ a bistability between states where both modes are locked. In this section, we show that this is possible using dual optical injection, a scheme that has been used to add functionality to multimode FP lasers\cite{horer_1997,jeong_2006}. In our model, dual injection is accounted for by allowing $e_m^{\mbox{\tiny inj}} \neq 0$. 

\begin{figure}
 \begin{center}
  \includegraphics[width=\columnwidth]{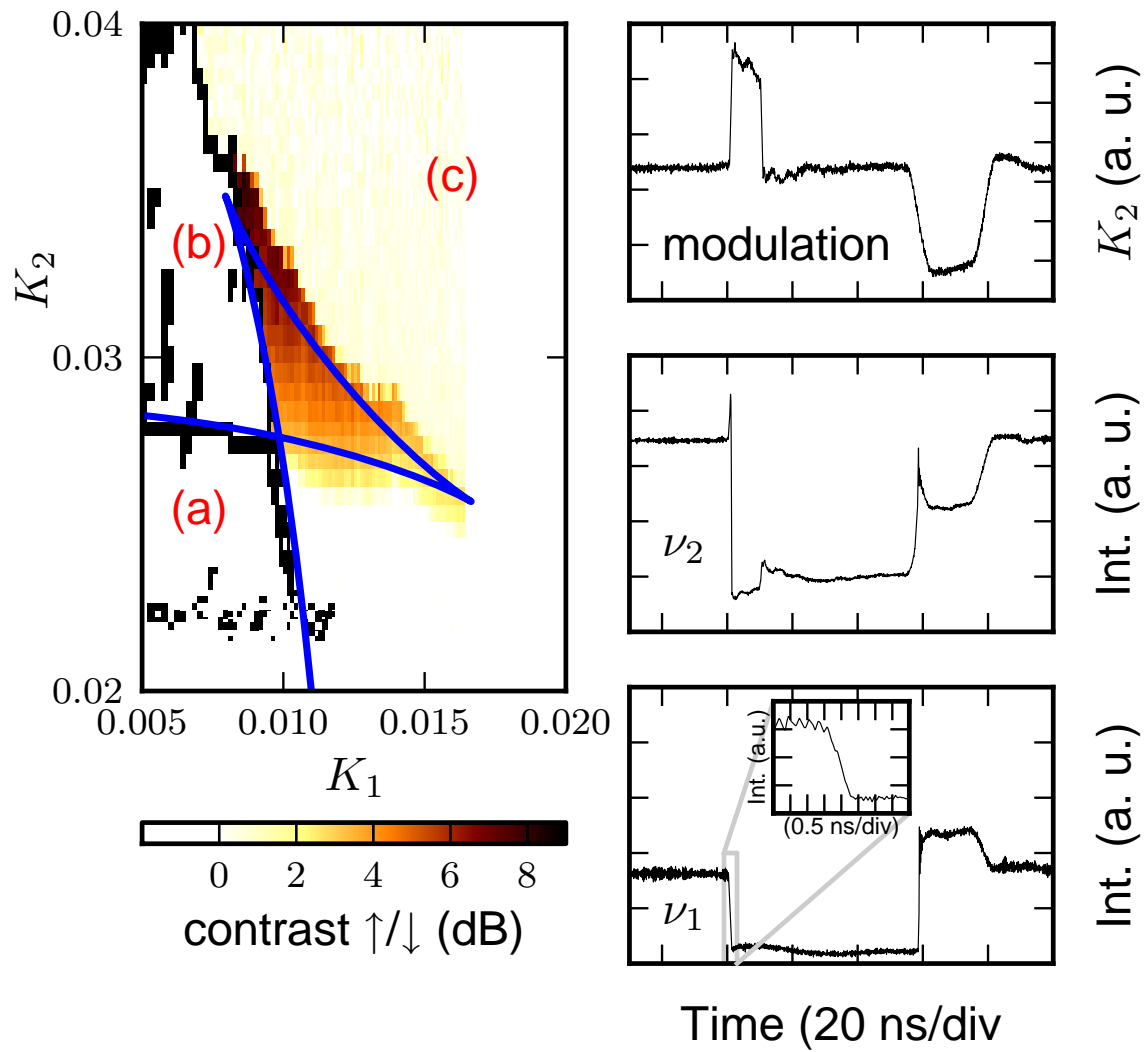}
   \caption{\label{fig:dual_smsr} The line in the left panel shows the calculated locking boundary for $\Delta\nu_2=-18\mbox{GHz}$ 
and $\Delta\nu_1=-8\mbox{GHz}$ with a triangular region indicative of bistability. The $K_1-K_2$ phase space is split into three regions: (a) where the laser in unlocked, (b) a bistability between locked and unlocked states, (c) the laser is locked. The color plot in the left panel shows the contrast of mode $\nu_1$ in a locked state (c) and shows the triangular bistability. The right panels show the experimental time series of $\nu_2$ and $\nu_1$ for the modulation shown.}
  \end{center}
\end{figure}

In the dual-injected system for equal detunings the locking boundary in the $K_1\mbox{-}K_2$ phase space is almost circular and by 
adjusting the $K_1$ and $K_2$ in the locking region power can be transferred continuously from one mode to the other. We have shown that for unequal detunings a bistability of these locked states arises \cite{heinricht_2001}. For the detunings $\Delta\nu_2=-18\mbox{ GHz}$ and $\Delta\nu_1=-8\mbox{ GHz}$ we show in figure \ref{fig:dual_smsr} the calculated boundary to the locking region (solid line), which forms a triangular region. The bistability arises inside this triangular region where two stable locked states exist. To demonstrate this bistability experimentally, we sweep the injection $K_2$ upward and downward for fixed $K_1$ and measure the contrast in optical power of mode $\nu_1$. We also monitor the r.f. spectra to determine whether the laser is injection locked or not and in this way we have identified three regions of dynamics: (a) no locked states, (b) bistability between a locked and unlocked state and (c) where the laser is always locked. Figure \ref{fig:dual_smsr} shows these three regions and the measured contrast of $\nu_1$ in the locked region (c) and we see the boundaries of non zero contrast is in excellent agreement with the calculated bifurcation diagram \cite{heinricht_2011}.

This bistability can also be exploited as an all-optical memory. For fixed $K_1$ and using same modulation of $K_2$ as before, the device acts as a memory element as shown in the left panels of figure \ref{fig:dual_smsr}. The switching time between states is now reduced to $\sim$ 500 ps which we attribute to the fact that the laser is always subject to strong injection during the switching process. However, due to the presence of the second holding beam the contrast has been reduced to $\sim$10 dB.

\subsection{Time delayed optical feedback}

The two previous schemes for an all-optical memory element employ holding beams which may not be ideal for practical systems. Not only will these beams increase the complexity of the system in general, they will also inevitably increase total power consumption, which is a major disadvantage also. An ideal memory element would be self-sustained and combine symmetry of the memory states with fast switching speed and high contrast. We have found that a memory element based on a two-colour device with delayed optical feedback can in principle satisfy all of these requirements. In this system bistability of single mode states is possible with switching enabled by injection of optical pulses into the appropriate wavelength
channel \cite{brandonisio_2012}. 

\begin{figure}
 \begin{center}
  \includegraphics[width=\columnwidth]{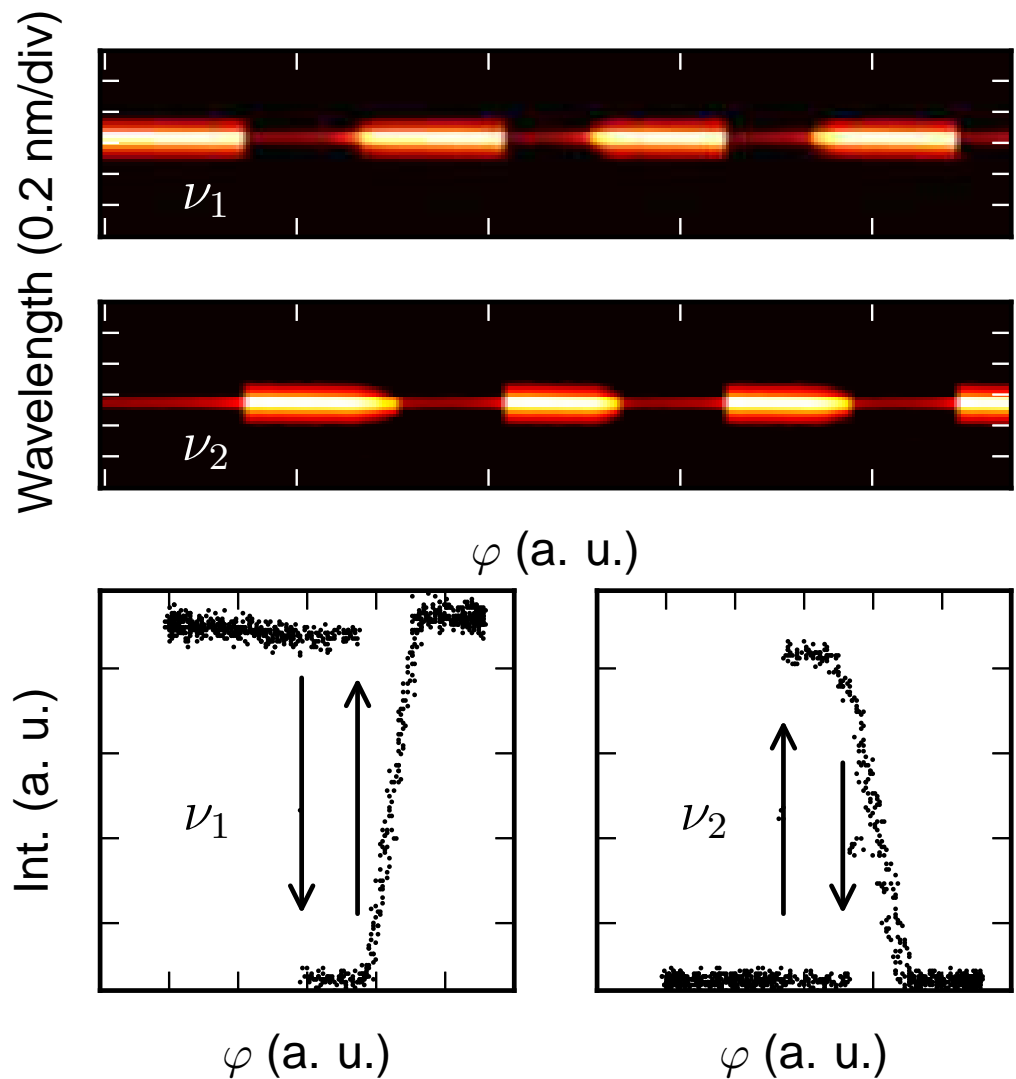}
   \caption{\label{fig:feedback} The top panels show the optical spectra of a two-colour laser with varying feedback phase around a 4 cm long cavity. Switching between the modes is seen with two distinct transitions. The bottom panels show the experimental modal intensities of $\nu_2$ and $\nu_1$ for a feedback phase modulated around a 1 cm cavity with the hysteresis clearly shown.}
  \end{center}
\end{figure}

For short external cavity lengths, small (wavelength scale) changes to the external cavity length can cause switching between the modes of the feedback system. The modes of the system in this case are the external cavity modes [ECMs], found from the solutions of the dual-mode Lang-Kobayashi equations \cite{lang_1980, heil_2003}. Experimentally, we use a semi-transparent mirror mounted on a piezo-electric translation stage to form a short external cavity. In figure \ref{fig:feedback}, wavelength switching is illustrated in the optical spectra for a two-colour laser with optical feedback using a 4 cm external cavity. The feedback strength in this case led to a 1\% reduction of threshold current and the full scale displacement of the feedback mirror is $\varphi \approx 3 \mu m$. Wavelength switching is clearly seen and two distinct switching behaviours are identified: (1) a smooth transition, $\nu_2$ to $\nu_1$ and (2) an abrupt transition from $\nu_1$ to $\nu_2$. The abrupt transistion is indicative of bistability and by using the piezo-electric translation stage to modulate the feedback phase around this abrupt transition, hysteresis can be demonstrated. Figure \ref{fig:feedback} shows the measured modal intensities which contains both the smooth and abrupt wavelength switches for a small modulation around a 1 cm long external cavity. The arrows indicate the direction of the hysteresis loop.

\begin{figure}
 \begin{center}
  \includegraphics[width=\columnwidth]{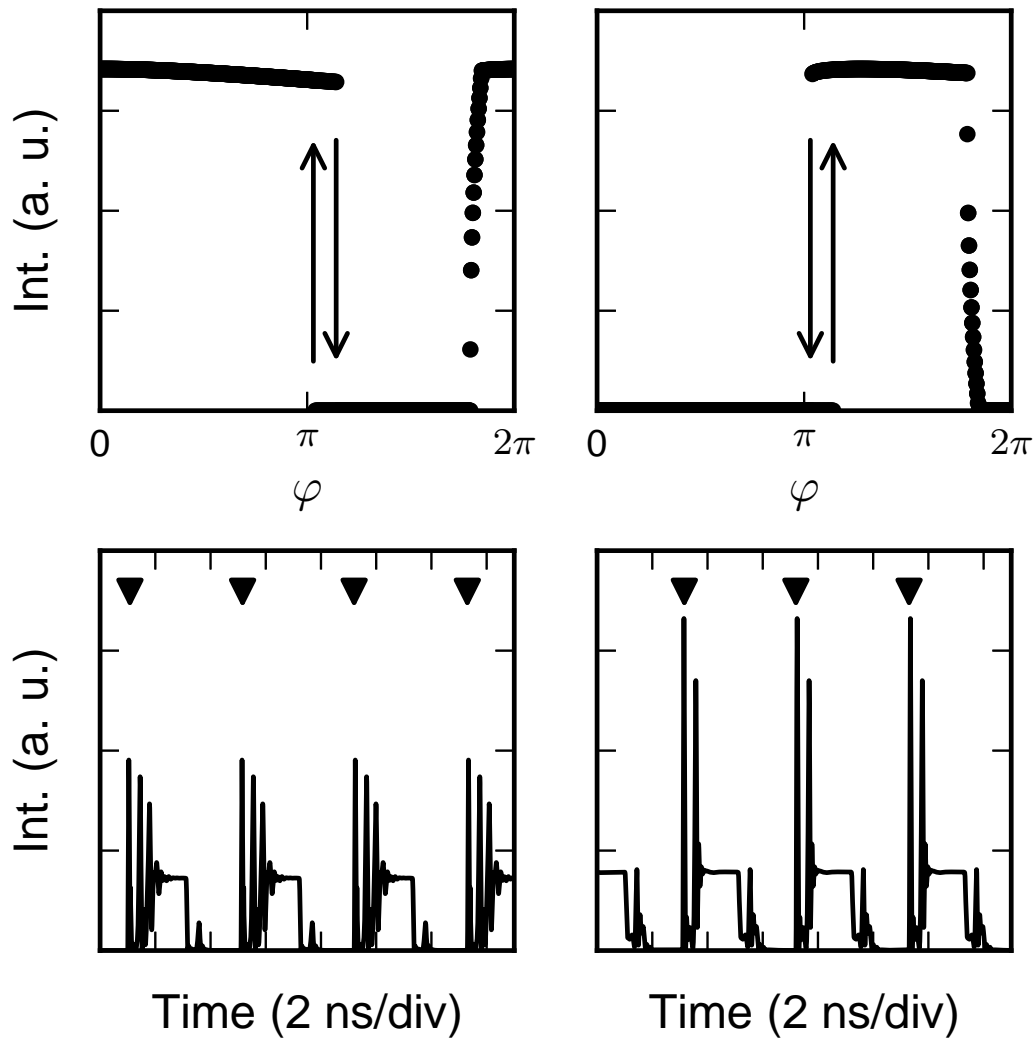}
   \caption{\label{fig:feedback_mem} The top panels show the calculated modal intensities of $\nu_2$ and $\nu_1$ for varying feedback phase around a 1 cm cavity and is in good agreement with the experiment. The lower panels show the switching dynamics of modal intensities with injected 100 ps pulses, which are indicated with the arrows.}
  \end{center}
\end{figure}

The system described in equations \ref{eq:sys} also provide a very good description of the optical feedback system. In figure \ref{fig:feedback_mem}, we plot the modal intensities for varying feedback phase for a 1 cm external cavity, which is close to the experimental value of figure \ref{fig:feedback}. Our theoretical analysis has shown that this bistability is bounded by saddle-node and transcritical bifurcations and can be optimized using the feedback strength and phase. This bistability can be exploited for an all-optical memory element when optical pulses are injected into the system at the frequencies corresponding to the detunings of the ECMs. In figure \ref{fig:feedback_mem} we show the switching dynamics of the modal intensities with 100 ps pulses, where we have used $\Delta\nu_1 = -3.5 \mbox{ GHz}$, $\Delta\nu_2=0.5 \mbox{ GHz}$, $\eta=0.018$, and $K_1=K_2=0.02$. The switching performance can be improved further with shorter cavities and for a 1 mm long cavity switching at a frequency of 5 GHz with optical pulses of duration equal to 10 ps could be achieved. These results suggest that the performance of the switching due to the feedback-induced bistability in microcavity lasers can be to very fast with pulse energies of a few femtojoules.

\section{Conclusions}

Wavelength switching dynamics of two-colour semiconductor lasers subjected to optical injection and feedback have been presented. We have shown how single and dual optical injection can induce bistability and that these bistabilities can be exploited as the basis for all-optical memory elements using modulated optical injection. Our experimental results show very good agreement with the results of numerical simulations based on systems of rate equations. We also showed that optical feedback can induce a bistability of single-mode states in the two-colour laser. Numerical simulations suggest that this memory element may be switched at high speed using injected optical pulses and without the requirement for an external holding beam.

\ack
The authors thank Eblana Photonics for the preparation of sample devices. This work was supported by Science Foundation Ireland and by Enterprise Ireland.


\begin{thebibliography}{00}

\bibitem{coldren}
Coldren L A and Corzine S W 1995 ``Diode lasers and photonic
integrated circuits,'' New York: Wiley

\bibitem{corbett_1995}
Corbett B and McDonald D 1995 Electronics Letters \textbf{31} 2181

\bibitem{obrien_2005}
O'Brien S and O'Reilly E P 2005 Appl. Phys. Lett. 86 201101

\bibitem{tani_2005}
Tani M, Morikawa O, Matsuura S, and Hangyo M, 2005 Semicond. Sci. Technol. 2005 \textbf{20} 151

\bibitem{white_2002}
White I, Penty R, Webster M, Chai Y J, Wonfor A, and Shahkooh S 2002 IEEE Communications Magazine 2002 \textbf{40} pp. 74-81 

\bibitem{brandonisio_2011}
Brandonisio N, Heinricht P, Osborne P, Amann A, and O'Brien S 2011 Journal of Optics \textbf{13} 125501

\bibitem{papadimitriou_2003}
Papadimitriou G I, Papazoglou C, Pomportsis A 2003 J. Lightwave Technology \textbf{21} 384 

\bibitem{dorren_2003}
Dorren H J S, Hill M T, Liu Y, Calabretta N, Srivatsa A, Huijskens F M, de Waardt H, and  Khoe G D J. 2003 Lightwave Technology \textbf{21} pp 2-12

\bibitem{osborne_2007}
Osborne S, O'Brien S, Buckley K, Fehse R, Amann A, Patchell J, Kelly B, Jones D E, O'Gorman J and O'Reilly E P 2007 IEEE J. Sel. Topics Quantum Electron \textbf{13} 1157 

\bibitem{obrien_2010}
O'Brien S, Osborne S, Bitauld D, Brandonisio N, Amann A, Phelan R, Kelly B and O'Gorman J 2010 IEEE Trans. Microwave Theory Tech. \textbf{58} 3083

\bibitem{yacomotti_2004} 
A. M. Yacomotti, L. Furfaro, X. Hachair, F. Pedaci, M. Giudici, J. Tredicce, J. Javaloyes and S. Balle, E. A. Viktorov and P. Mandel Phys. Rev. A 2004 \textbf{69}, 053816.

\bibitem{Liu_2010}
Liu L, Kumar R, Huybrechts K, Spuesens T, Roelkens G, Geluk E-J, de Vries T, Regreny P, Van Thourhout D, Baets R and Morthier G 2010 Nature Photonics \textbf{4} pp182-187

\bibitem{hill_2004}
Hill M T, Dorren H J S, de Vries T, Leijtens X J M, den Besten J H, Smalbrugge B, Oei Y, Binsma H, Khoe G, and Smit M K 2004 Nature \textbf{432}, pp 206-209

\bibitem{mori_2006}
Mori T, Yamayoshi Y, and Kawaguchi H 2006 Appl. Phys. Lett. \textbf{88} 101102

\bibitem{zhukovsky_2009}
Zhukovsky S V, and  Chigrin D N 2009 Optics Letters \textbf{34} pp 3310-3312

\bibitem{chen_2011}
Chen C H, Matsuo S, Nozaki K, Shinya A, Sato T, Kawaguchi Y, Sumikura H, and Notomi M 2011 Optics Express \textbf{19} pp 3387-3395

\bibitem{osborne_2009}
Osborne S, Buckley K, Amann A, and O'Brien S 2009 Optics Express \textbf{17} pp 6293-6300

\bibitem{varangis_1997}
Varangis P M, Gavrielides A, Erneux T, Kovanis V, and Lester L F 1997 Phys. Rev. Lett. \textbf{78} pp2353-2356

\bibitem{osborne_2009_PRA}
Osborne S, Amann A, Buckley K, Ryan G, Hegarty S P, Huyet G, and O'Brien S, 2009  Phys. Rev. A \textbf{79} 023834

\bibitem{brandonisio_2012}
Brandonisio N, Heinricht P, Osborne P, Amann A, and O'Brien S 2012 IEEE Photonics Journ. \textbf{4} 95

\bibitem{auto}
Doedel E J \textit{et al}. 
Technical report 2007 Concordia University Montreal [http://indy.cs.concordia.ca/auto/]

\bibitem{gatare_2009}
Gatare I, Sciamanna M, Nizette N, Thienpont H, and Panajotov K 2009 Phys. Rev. E \textbf{80}, 026218  

\bibitem{coomans_2010}
Coomans W, Beri S, Van der Sande G, Gelens L, and Danckaert J 2010 Phys. Rev. A \textbf{81}, 033802

\bibitem{horer_1997}
H\"orer J and Patzak E 1997 IEEE J. Quantum Electron. \textbf{33} 596

\bibitem{jeong_2006}
Jeong Y D, Cho J S, Won Y H, Lee H J, and Yoo H 2006 Opt. Express \textbf{14} 4058

\bibitem{heinricht_2011}
Heinricht P, Wetzel B, O'Brien S, Amann A and Osborne S 2011 Appl. Phys. Lett. \textbf{99} 011104

\bibitem{lang_1980}
Lang R, and Kobayashi K, 1980 IEEE J. Quantum Electron. \textbf{16} pp 347-355

\bibitem {heil_2003}
Heil T, Fischer I, Elsasser W, Krauskopf B, Green K, and Gavrielides A 2003 Phys. Rev. E, \textbf{67} 066214

\end{thebibliography}
\end{document}